\documentclass[10pt, conference]{IEEEtran}
\IEEEoverridecommandlockouts

\usepackage{cite}
\usepackage{amsmath,amssymb,amsfonts}
\usepackage{algorithmic}
\usepackage{graphicx}
\usepackage{makecell}
\usepackage{textcomp}
\def\BibTeX{{\rm B\kern-.05em{\sc i\kern-.025em b}\kern-.08em
    T\kern-.1667em\lower.7ex\hbox{E}\kern-.125em}}

\usepackage{xcolor}
\usepackage{float}

\begin{document}

\title{AICons: An AI-Enabled Consensus Algorithm Driven by Energy Preservation and Fairness}

\author{
\IEEEauthorblockN{
Qi Xiong\IEEEauthorrefmark{1} \hspace*{.15in}
Nasrin Sohrabi\IEEEauthorrefmark{2} \hspace*{.15in}
Hai Dong\IEEEauthorrefmark{2} \hspace*{.15in}
Chenhao Xu\IEEEauthorrefmark{2}  \hspace*{.15in}
Zahir Tari\IEEEauthorrefmark{2}
}

\IEEEauthorblockA{
\IEEEauthorrefmark{1}\IEEEauthorrefmark{2}School of Computing Technologies \\ Centre of Cyber Security Research \& Innovation (CCSRI) \\ RMIT University, Melbourne, Australia\\
Contact person: 
\IEEEauthorrefmark{1} \it qi.xiong@student.rmit.edu.au}
}

\maketitle

\begin{abstract}
Blockchain has been used in several domains. However, this technology still has major limitations that are largely related to one of its core components, namely the consensus protocol/algorithm. Several solutions have been proposed in literature and some of them are based on the use of Machine Learning (ML) methods. The ML-based consensus algorithms usually waste the work done by the (contributing/participating) nodes, as only winners' ML models are considered/used, resulting in low energy efficiency. To reduce energy waste and improve scalability, this paper proposes an AI-enabled consensus algorithm (named AICons) driven by energy preservation and fairness of rewarding nodes based on their contribution. In particular, the local ML models trained by all nodes are utilised to generate a global ML model for selecting winners, which reduces energy waste. Considering the fairness of the rewards, we innovatively designed a utility function for the Shapley value evaluation equation to evaluate the contribution of each node from three aspects, namely ML model accuracy, energy consumption, and network bandwidth. The three aspects are combined into a single Shapley value to reflect the contribution of each node in a blockchain system. Extensive experiments were carried out to evaluate fairness, scalability, and profitability of the proposed solution. In particular, AICons has an evenly distributed reward-contribution ratio across nodes, handling 38.4 more transactions per second, and allowing nodes to get more profit to support a bigger network than the state-of-the-art schemes.
\end{abstract}


\section{Introduction}
Blockchain is increasingly used in several domains. Through the use of a consensus protocol/algorithm, a blockchain system allows its nodes to reach an agreement without a central authority. However, consensus algorithms also bring their own challenges when it comes to key aspects (e.g., energy efficiency and scalability), which limits their use in broader fields. For example, in Bitcoin, the Proof-of-Work (PoW) consensus algorithm consumes a lot of energy to select winners who are responsible for creating new blocks, resulting in very low scalability. Another well-known consensus algorithm is Proof-of-Stake (PoS), which improves energy efficiency and scalability. However, it introduces other issues such as centralization~\cite{saad_e-pos_2021} and long range attack~\cite{deirmentzoglou2019survey}.

With the advances of Machine Learning (ML) models, several attempts focused on their use to improve energy efficiency of blockchain systems~\cite{baldominos_coinai_2019, chenli_energy-recycling_2019, bravo-marquez_proof--learning_2019, lihu_proof_2020, lan_proof_2021}. In these  schemes, the consensus process requires nodes to compete for training ML models, where rewards are given to the node (winner) whose ML model first achieves the desired accuracy. They are called proof-of-Useful-Work (PoUW). However, the work done by other nodes (non-winners) is ``discarded", causing the computing resources to be wasted. 

To reduce energy wastage in blockchain systems, this paper proposes an AI-enabled consensus algorithm (AICons) to reduce energy consumption as well as to improve scalability. In particular, based on the monitoring of winners' data in a blockchain system, all nodes train their ML-based recommender system model~\cite{portugal2018use} that recommends a winner to the blockchain system. The data monitored during this process including CPU, memory, and network bandwidth of the participating nodes are collected by using a secure shell protocol (SSH). Under the decentralised federated learning framework~\cite{hu2019decentralized}, each node collects and aggregates all local ML models into a unified global ML model, which is later used to select a winner based on the current monitored data. AICons makes the work of all nodes as useful work, therefore improving the energy efficiency of blockchain systems. Besides, the aggregated global model mitigates the malicious ranking behaviours of nodes while improving scalability.

Existing blockchain systems reward nodes that optimise only a single parameter, such as {\it stakes} in PoS, or {\it storage space} in Proof-of-Space~\cite{ateniese_proofs_2014}. To evaluate the contribution of nodes more comprehensively, we use the Shapley value function~\cite{shapley1997value} in blockchain systems by treating the ML model training process as a cooperative game. Specifically, we creatively design a utility function that optimises the contribution of nodes from three parameters, namely the {\it ML model accuracy}, {\it energy consumption}, and {\it network bandwidth}. The use of these three parameters together further reduces the energy consumed by the nodes while improving scalability.

To the best of our knowledge, this is the first attempt to design a consensus algorithm that makes utilisation of the works done by all nodes and therefore improves energy efficiency. The design of a multi-dimensional contribution evaluation mechanism for blockchain systems using Shapley value is also an innovative approach to integrate various parameters (e.g., energy, bandwidth) to improve existing blockchain systems. The contributions of this paper can be summarised as follows.
\begin{itemize}
    \item An AI-enabled consensus algorithm (AICons) based on decentralised federated learning that allows nodes to collaboratively train a recommender system model (which is an ML model) for selecting winners. Specifically, nodes train their local ML models and broadcast them to other nodes. When broadcasted ML models are received by nodes, they merge them into a global ML model. The global ML model is then used for recommending winners to the blockchain system.
   
    \item A fair rewarding mechanism that further motivates nodes to reduce energy consumption while improving scalability by creatively introducing a utility function to the Shapley value. This function measures the contribution of nodes from three dimensions, (i.e.,  ML model accuracy, energy consumption, and network bandwidth). Higher model accuracy, lower energy consumption, and higher bandwidth mean higher contribution. The three-dimensional metrics are combined into a single Shapley value as the reward to nodes for their contributions to blockchain. 
    
\end{itemize}

We conducted experiments where AICons is benchmarked against four state-of-the-art consensus algorithms with metrics related to fairness, scalability, reward and profitability. The experiment results show that, compared with the state-of-the-art consensus algorithms, AICons is fairer in rewarding nodes as it has an evenly distributed reward-contribution ratio across nodes. AICons is also more scalable when the number of nodes increases, as it processes 38.4 more transactions per second than others. Finally, AICons allows nodes to get more profit and thereby supports bigger networks.

The rest of the paper is organised as follows. Section~\ref{sec:related_work} summarises some of the existing works, and Section~\ref{sec:background} explains some of the basics of Shapely value. Section~\ref{sec:solution} provides details of AICons, and Section~\ref{sec:evaluation} discusses the various experimental results. Concluding remarks are provided  in Section~\ref{sec:conclusion}.

\section{Related Work}
\label{sec:related_work}

This section firstly describes the early consensus algorithms for blockchain systems (called here ``traditional" consensus algorithms) and then consensus algorithms that integrate  ML.

\subsection{``Traditional" Consensus Algorithms}
Bitcoin uses PoW as the consensus algorithm to select winners who are responsible for creating new blocks in blockchain systems. PoW's nodes compete to find a hash number that satisfies the requirement of the blockchain system. The first one who finds the hash number is selected as the winner. However, the high energy consumption induced in finding the hash number is widely criticised~\cite{saad_e-pos_2021}. This issue is addressed by an alternative consensus algorithm, called PoS (Proof of Stake), which eliminates the computing power competition by selecting winners based on the stakes of nodes. The blockchain system selects a coin from all available coins and the owner of the selected coin is the winner. Hence those who have the most coins have a higher chance of being selected as the winner. However, PoS introduces security concerns, including centralization~\cite{saad_e-pos_2021} and long-range attacks~\cite{deirmentzoglou2019survey}, because rich nodes have a higher probability of being selected and becoming richer. Byzantine Fault Tolerance (BFT) protocols like HoneyBadgerBFT~\cite{miller_honey_2016} in which nodes broadcast the transactions they select from their local cache to other nodes and vote on the received transactions to decide whether the transactions they receive should be included in the new block or not. When over 2/3 of votes are positive, then the transactions are included in the newly created block. It does not have issues of high power consumption or centralisation, but the lack of incentive mechanisms resulting in them being only suitable for private blockchains.

Proof of Reputation (PoR)~\cite{gai_proof_2018}, Proof of Interest (PoI)~\cite{new_economy_movement}, Proof of Devotion (PoD)~\cite{nebulas_whitepaper}, and Proof of Believability (PoB)~\cite{noauthor_iost_nodate} use a scoring mechanism to evaluate the quality of nodes based on their past behaviours (e.g., whether they cheated or not) and select the one who has the highest score as the winner. PoR uses a global reputation system that ranks nodes based on their credit score. The node with the highest credit score creates a new block. Nodes are rated a credit score by providing services to other nodes. The rate is based on the quality of the service the node provided to other nodes. PoI measures the importance score of each node based on their accumulated contribution to the blockchain system such as committing transactions that help flourish the blockchain community. The node who has the highest score has the highest probability of becoming a block proposer. PoD ranks nodes based on their devotion (e.g., the frequency of committing transactions, the number of connections with other nodes, frequent interactions with smart contracts and decentralised applications) in the blockchain community. The top $n$ ranked nodes are selected as validators that are responsible for creating new blocks. PoB divides nodes into two groups that are believable/normal leagues. The nodes from the believable group pack transactions into new blocks. The nodes from the normal league verify the transactions in the new blocks. The nodes are selected into the believable league based on their believability score which is calculated based on their contributions to the blockchain community such as token balance.

\subsection{ML-based Consensus Algorithms}
It is intuitive to use ML models in blockchain consensus algorithms to optimise energy consumption. Several PoUW schemes~\cite{baldominos_coinai_2019, chenli_energy-recycling_2019, bravo-marquez_proof--learning_2019, lihu_proof_2020, lan_proof_2021} were proposed to train ML models in a blockchain system. The ML models to be trained are published as training tasks by providers. The providers need to deposit tokens in advance, and trainers are rewarded for the training task. The consensus process is defined as an ML model training process, where only one node is rewarded and thus the works carried out by other nodes are wasted. Proof-of-Federated-Learning (PoFL)~\cite{qu_proof_2021} reduces the energy waste of non-winners by training the ML model in a federated manner. PoFL's nodes are divided into groups and each group trains their federated ML model. The group whose ML model accuracy is the highest wins and the reward is shared among the winning group. Compared with the aforementioned PoUW schemes, it reduces the waste works. However, the works of non-winner groups are still abandoned. By contrast, this paper proposes an AI-enabled consensus algorithm (AICons) that reduces energy waste on non-winners and guarantees fair rewards to all nodes based on their contribution.

\section{Background: Shapley Value}
\label{sec:background}
To improve energy efficiency, and scalability of blockchain systems, we use the Shapley value to fairly evaluate the contribution of nodes and issue rewards. The Shapley value~\cite{shapley1997value} is a solution used in game theory that aims to fairly distribute the game reward to participants in a cooperative game. It calculates a marginal contribution for each participant. The reward distribution is based on the marginal contribution. For example, the marginal contribution of a node, say $i$, is calculated by the Shapley value equation

\begin{equation}
    s_{i}=\frac{1}{\left | N \right |}\sum _{S\subseteq N\setminus \{i\}}\binom{\left | N \right |-1}{\left | S \right |}^{-1}U(S \cup i)-U(S),
    \label{eq:original_shapley_value_function}
\end{equation}

\noindent in which the node $i$ is combined with a permutation of subsets $S$ (within the set of all nodes $N$) of other nodes and measures the difference of the gain (which is computed by a utility function $U$) between with and without the existence of $i$. If the gain when the node $i$ is included is larger than when $i$ is not included,  $i$ is to said to positively contribute to the cooperative game.

\section{The Proposed AICons Solution}
\label{sec:solution}
AICons contains three parts: decentralised federated learning, AI-enabled node ranking, and fair contribution evaluation. 
Specifically, a decentralised federated learning architecture is adopted to allow nodes to train the ML model that is used for ranking nodes. The AI-enabled node ranking process utilises the ML models trained by nodes to generate the final rank for the nodes. The fair contribution evaluation mechanism computes the nodes' three dimensional contributions for fairly issuing rewards while motivating nodes to improve the quality of their ML models, reduce energy consumption, and provide more network bandwidth to the blockchain system.

\subsection{Decentralised Federated Learning}
To improve energy efficiency and scalability of blockchain systems, we used ML models to select winners in (blockchain) networks for generating new blocks. ML models may bring trust issues to public blockchain systems if they are trained and deployed by a single node, as bias can be introduced in ML models. Considering the security concern of the trust issues and the fact that a public unprivileged blockchain system is based on a decentralised architecture, we introduced a decentralised federated learning framework to train the ML model. Specifically, each node of the blockchain takes charge of training and broadcasting their local ML models, as well as aggregating the local ML models from other nodes. By doing so, the impact of malicious local ML models is mitigated in the global ML model.

According to~\cite{xu2021asynchronous}, classic federated learning requires three steps to finish one epoch of ML model training. 1) A central server initializes the parameters (learning rate, batch size, etc.) of the global model and broadcasts it to clients who participate in the federated training. 2) Clients train their local models based on the global model and upload their local model to the server once finished the training. 3) The server aggregates the local models into a new global model by using
\begin{equation}
    w_g=\frac{1}{n}\sum_{i=1}^n w_i,
    \label{eq:fl_agg}
\end{equation}
where $w_g$ is the global model, $w_i$ is the local model of node $i$.

\begin{figure}
    \centering
    \includegraphics[width=1\linewidth]{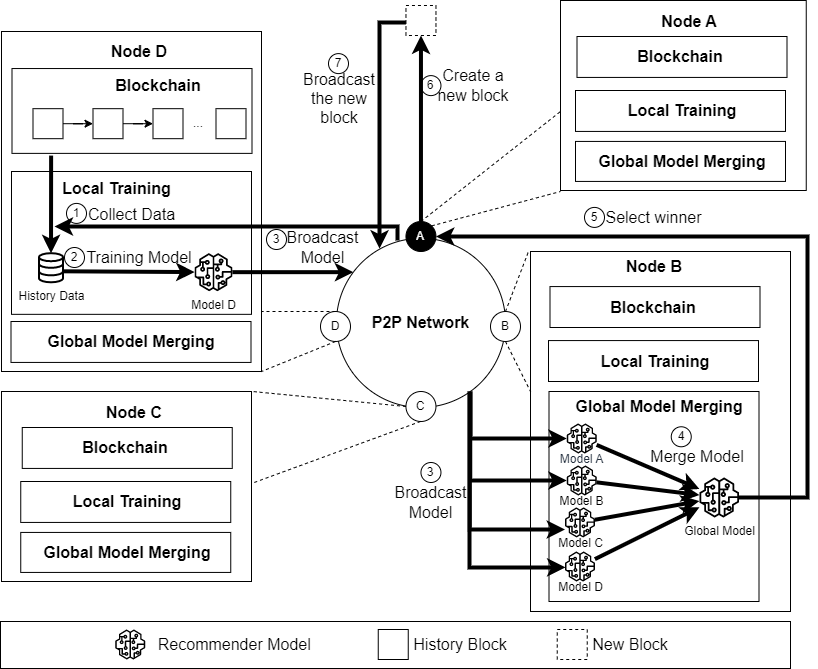}
    \caption{The architecture of decentralised federated learning and the workflow of ML model training.}
    \label{fig:arch_federate_learning}
\end{figure}

In decentralised federated learning, the global model aggregation is similar to classic federated learning, however, no central server is needed. The workflow of the decentralised federated learning in AICon is shown in Figure~\ref{fig:arch_federate_learning}. There are four nodes, including Node A, Node B, Node C, and Node D, participating in the blockchain network and communicating with each other through a P2P network. All nodes in the blockchain network have a replica of the shared ledger and two functions, which are Local Training and Global Model Merging. 

The process of decentralised federated learning in AICons includes seven steps, as shown below. 
\begin{itemize}
\item {\bf Step 1}: Every node collects historical data from the blockchain as well as real-time monitoring data from other nodes, which includes the data of previous winners, which is used as labelled data for training the ML model.

\item  \textbf{Step 2}. Every node trains its local ML model by using the data it collected from Step 1 as the input to the ML model. The ML model works as a recommender system that suggests the top $n$ winner candidates in the blockchain network. 

\item \textbf{Step 3}: After the training, every node broadcasts its trained local ML model to the blockchain network. 

\textbf{Step 4}: When a node receives local ML models broadcasted by other nodes, it merges them into a global ML model. This is done in a similar way to classical federated learning~\cite{xu2021asynchronous}.

\item \textbf{Step 5}: Every node runs the global ML model and takes the real-time monitoring data collected from other nodes as input. The output of the ML model is a rank list of nodes, which is then utilised for winner selection. The detailed AI-enabled node ranking process is explained in Section~\ref{sec:node_ranking}.

\item \textbf{Step 6}: If a node finds itself selected as the winner, it creates a new block. The block contains transactions and the monitoring data of the current winner. The winner also writes the currently trained ML model into the block so that newly jointed nodes can fetch it from the blockchain.

\item \textbf{Step 7}: The winner broadcast the newly created block to the blockchain network.
\end{itemize}

\subsection{AI-enabled Node Ranking}
\label{sec:node_ranking}

The AI-enabled node ranking in AICons is the process that utilises the ML models trained by decentralised federated learning to rank nodes as well as to recommend winners for generating a new block. The AI-enabled node ranking process helps nodes to achieve consensus on the final rank of nodes. By adopting the ML model to rank nodes and select the winner, the consensus process is ensured to be energy efficient and scalable. The notation used in this section is listed in Table~\ref{tab:notation_federated_learning}. 

\begin{table}[htbp]
    \renewcommand{\arraystretch}{1.3}
    \caption{Notation for Recommender System Model}
    \centering
    \begin{tabular}{c|c}
    \hline
    \textbf{Symbol} & \textbf{Definition} \\
    \hline
    $l$ & Loss function \\
    \hline
    $f_s$ & Cosine similarity function \\
    \hline
    $z_i$ & Embedding of node $i$ \\
    \hline
    \end{tabular}
    \label{tab:notation_federated_learning}
\end{table}

\begin{figure}[htbp]
    \centering
    \includegraphics[width=1\linewidth]{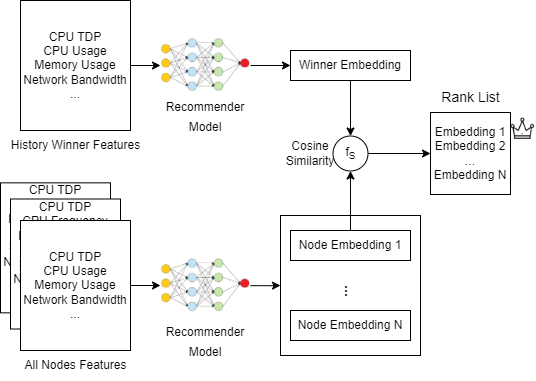}
    \caption{The AI-enabled node ranking and winner selection.}
    \label{fig:ai_consensus _model}
\end{figure}

The AI-enabled node ranking process is shown in Figure~\ref{fig:ai_consensus _model}. The ML model (recommender system model) embeds all nodes into a vector space in which all nodes are represented as an embedding~\cite{collobert_unified_2008}. The node embedding list is ranked by the similarity to the expected winner (we use history winners as the expected winner). The one with the highest similarity score is recommended as the winner. 

To guarantee energy efficiency and scalability of blockchain systems, the monitoring data contains information about CPU TDP, CPU usage, memory usage, and network bandwidth.

To maximise the margin between true positive and false positive metrics (the metric is the cosine similarity between two nodes in this case), we use a margin-based loss function~\cite{ying_graph_2018}
\begin{equation}
l=\frac{1}{n}\sum_{i=0}^{n}max(0, \frac{1}{m}\sum_{j=0}^{m}f_c(\textbf{z}_i, \textbf{z}_{n_j})+\Delta-\frac{1}{h}\sum_{j=0}^{h}f_c(\textbf{z}_i, \textbf{z}_{p_j})),
\label{loss_function}
\end{equation}
\noindent where $\textbf{z}_i$ is the embedding vector of the node $i$ drawn from the set $N$. $\textbf{z}_{n_j}$ represents the negative node (sampled from negative node set), which is not similar to the node $i$. $\textbf{z}_{p_j}$ represents the positive node (sampled from the positive set) which is similar to the node $i$. $\Delta$ is the margin hyper-parameter \cite{ying_graph_2018}. The similarity of the embedding vectors between two nodes is calculated by a cosine similarity function:
\begin{equation}
f_c(\textbf{z}_i,\textbf{z}_j)=\|\textbf{z}_i\|\|\textbf{z}_j\|cos(\theta)=\frac{\textbf{z}_i\cdot\textbf{z}_j}{\|\textbf{z}_i\|\|\textbf{z}_j\|},
\label{cos_sim}
\end{equation}
\noindent where $\textbf{z}_i$ is the embedding vector of node $i$ and $\textbf{z}_j$ is the embedding vector of node $j$. If the cosine similarity between node $i$ and node $j$ is close to 1, and if between node $i$ and $k$ is close to 0 (i.e., the opposite of what we want), then the loss $l$ will be very large. The neural network needs to tune its weight to reduce the loss so that the cosine similarity between  $i$ and  $j$ is close to 0, and the cosine similarity between nodes $i$ and $k$ is close to 1.

\subsection{Fair Contribution Evaluation}

\begin{figure*}
  \centering
  \includegraphics[width=1\linewidth]{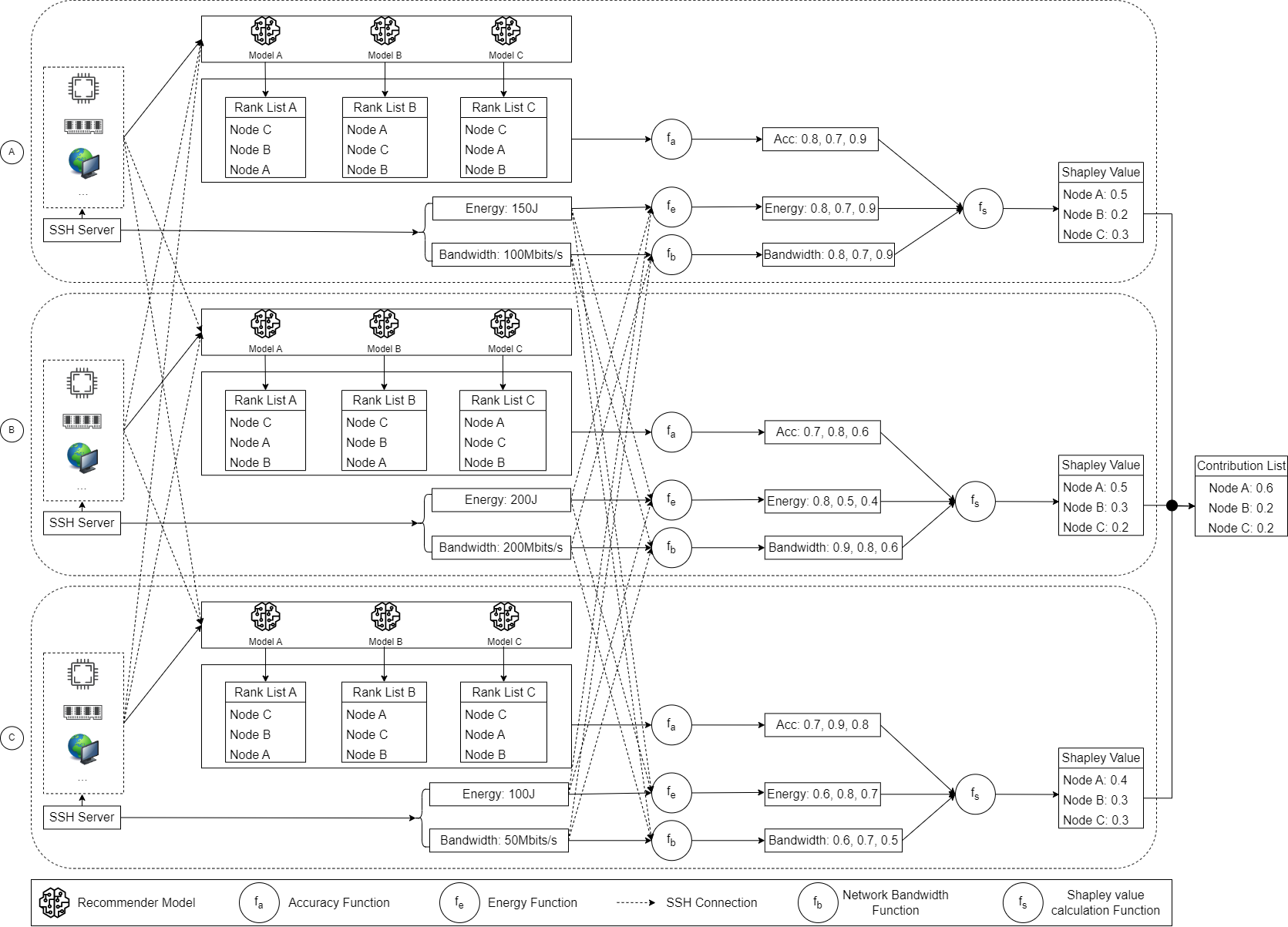}
  \caption{Calculate the contribution of Node A, B, and C.}
  \label{fig:arch_shapley_value}
\end{figure*}

To fairly evaluate the contribution of nodes and reward them accordingly, we use the Shapley value in the proposed reward scheme. Specifically, the Shapley value~\cite{shapley1997value} is used to measure the contributions of nodes in cooperative games. Since the participation of training the ML model can be modeled as a cooperative game, and the Shapley value guarantees the fairness of the measured contribution, we use the Shapley value to measure the contribution of each node. The notation used in this section is listed in Table~\ref{tab:notation_shapely_value}.

\begin{table}[htbp]
    \renewcommand{\arraystretch}{1.3}
    \caption{Notation for the Shapley Value Equations}
    \centering
    \begin{tabular}{c|p{0.76\linewidth}}
    \hline
    \textbf{Symbol} & \textbf{Definition} \\
    \hline
    $\textbf{s}_{i,j}$ & Shapley value vector of node $i$ calculated by node $j$ \\
    \hline
    $f_c$ & Cosine similarity function \\
    \hline
    $\phi$ & Marginal contribution calculation function \\
    \hline
    $f_a$ & Calculation function of ML model accuracy \\
    \hline
    $f_s$ & Calculation function of Shapley value \\
    \hline
    $f_e$ & Calculation function of contribution in energy consumption \\
    \hline
    $f_b$ & Calculation function of contribution in network bandwidth \\
    \hline
    \end{tabular}
    \label{tab:notation_shapely_value}
\end{table}

To incentivise nodes to improve the quality of the ML model while reducing energy consumption, and to also maximise the total network bandwidth in the blockchain system, we designed a utility function for the Shapley value that {\bf considers three parameters}, namely ML model accuracy, energy consumption, and network bandwidth. The process of calculating the Shapley value (contribution) is illustrated in Figure~\ref{fig:arch_shapley_value}. We used here three nodes (i.e., Node A, Node B, and Node C) to illustrate how the Shapley value is computed. 

After training the ML model, each node has all the local ML models (broadcasted by other nodes). They then merge all the local ML models into a global ML model. Each node takes the monitoring data collected from other nodes as the input of the global ML model. The output of the global ML model is a top $n$ ranked list. 

To avoid the situation in which some nodes may send fake data to other nodes, each node uses an SSH client to connect to an SSH server hosted by other nodes (i.e., SSH host) and collects the energy consumption and network bandwidth of other nodes in real time. Each node runs a command line (e.g., a resource monitoring tool on other nodes' local environment) to collect the aforementioned data. Then, each node takes accuracy, energy consumption, and network bandwidth information as the input to a Shapley value calculating function and gets the output of a list of Shapley values (that indicate the contribution of every node). After computing the Shapley value, each node broadcasts the results to other nodes. When receiving all Shapley values broadcasted by others, each node merges the Shapley values into a contribution list. The contribution list is then used to issue rewards to nodes.

The Shapley value equation~\cite{shapley1997value} is defined as
\begin{equation}
    \textbf{s}_{i,j}=\frac{1}{\left | N \right |}\sum _{S\subseteq N\setminus \{i\}}\binom{\left | N \right |-1}{\left | S \right |}^{-1}\phi(S,i),
    \label{eq:shapley_value_function}
\end{equation}
\noindent where $N$ is the whole set of nodes; $S$ is the subset of $N$; $\phi$ is a function that computes the marginal contribution of the node $i$ in the subset $S$. $s_{i,j}$ is the Shapley value of the node $i$ computed by the node $j$. The marginal contribution of the node $i$ is carried out by calculating the difference between the subset with the node $i$ and the subset without the node $i$. The subsets are from the permutation of the whole set $N$. According to~\cite{shapley1997value}, the marginal contribution of node $i$ is the difference between the utility with node $i$ and the utility without node $i$. Therefore, the function $\phi$ in (\ref{eq:shapley_value_function}) is defined as
\begin{equation}
    \phi(S,i) = U(W_{S\cup\{i\}}, E_{S\cup\{i\}}, B_{S\cup\{i\}})-U(W_{S}, E_{S}, B_{S}),
    \label{eq:shapley_value_phi}
\end{equation}
\noindent where $W_{S}$ is the weight parameters of a subset of nodes $S$; $E_{S}$ is the energy consumption of the subset of nodes $S$; $B_{S}$ is the total network bandwidth of $S$. $W_{S}$, $E_{S}$, and $B_{S}$ are input parameters of the utility function $U$. Formally, we define $W_{S}$, $E_{S}$, $B_{S}$ in (\ref{eq:shapley_value_phi}) as
\begin{align}
W_S &= \{w_k|k\in S\}; \nonumber \\
E_S &= \{e_k|k\in S\}; \nonumber \\
B_S &= \{b_k|k\in S\}.
\label{eq:permutation_without_i}
\end{align}

To include three parameters in the Shapley value, we have introduced the utility function $U$ in (\ref{eq:shapley_value_phi}) as a function that takes three inputs of aforementioned $W_{S}$, $E_{S}$, $B_{S}$ and outputs a vector with three elements (i.e., the three parameters we considered. We use elements and parameters interchangeably throughout the paper):

\begin{align}
U(W_{S}, E_{S}, B_{S}) &= [f_a(W_{S}), f_e(E_{S}), f_d(B_{S})] \nonumber \\
&= [a, e, b],
\label{eq:utility_function}
\end{align}
\noindent where the three inputs are: the ML model parameters $W_{S}$, the energy consumption $E_{S}$, and the network bandwidth $B_{S}$. They are collected from the subset $S$ or $S\cup \{i\}$. The output is a vector with three elements that contains the subset's overall accuracy $a$, energy consumption $e$, and network bandwidth $b$. 

To calculate the three elements of the output vector, we introduce three sub-functions $f_{a}$, $f_{e}$, and $f_{b}$ in (\ref{eq:utility_function}). Specifically, for $f_{a}$, we have 
\begin{equation}
f_{a}(W_{S})=f(y=\hat{y}|\textbf{x}, Agg(W_{S})),
\label{eq:fun_a}
\end{equation}
\noindent that computes the {\it overall accuracy} of the subset. Since the ML model is trained together by nodes through federated training, the function $f_a$ needs firstly to aggregate the ML model parameters $\textbf{w}$ by using the $Agg$ function, and then calculate the accuracy of the aggregated version of the ML model. To compute the accuracy, the test data $\textbf{x}$ is taken as the input of the aggregated ML model, and then the accuracy is computed by testing whether the predicted output $\hat{y}$ is equal to the real output $y$. For $f_{e}$, we have
\begin{equation}
f_{e}(E_{S})=\sum_{i=0}^{|S|}\frac{1}{e_i},
\label{eq:fun_e}
\end{equation}
\noindent to calculate the {\it total energy consumed} by the subset $S$. Since those who consumed less energy contributed more to saving energy, we designed the $f_e$ function as the sum of the reciprocal of the energy consumption of each node from the subset $S$. For $f_{b}$, we have
\begin{equation}
f_b(B_{S})=\sum_{i=0}^{|S|}b_i,
\label{eq:fun_d}
\end{equation}
\noindent to calculate the {\it total network bandwidth} of the subset $S$. Since those who contributed more network bandwidth contribute more to the total network bandwidth, the function $f_d$ computes the total sum of the network bandwidth of each node.

To measure the energy consumption of each node in (\ref{eq:fun_e}), we have 
\begin{equation}
    e_{i} = p^{cpu}\times t^{cpu},
    \label{eq:vector_ele_e}
\end{equation}
\noindent that multiplies the CPU TDP $p^{cpu}$ and the running time $t^{cpu}$.

To quantify the network status, namely $b_{i}$ in (\ref{eq:fun_d}), this is defined as the network bandwidth between the node $i$ and other nodes. The communication between nodes is frequent in real scenarios, therefore one can use the average bandwidth as the expected network bandwidth to better measure the network bandwidth.

After the node $j$ calculates the Shapley value vector for each node by using the above equations, we group the Shapley value vectors $\textbf{s}_i$. For simplicity in the calculation, we can stack those vectors vertically to get the node $j$'s Shapley value matrix ($R_{|N|\times 3}^j$ means the Shapley value matrix calculated by the node $j$) as
\begin{equation}
    R_{|N|\times 3}^j=
    \begin{bmatrix}
        \textbf{s}_{1,j}\\
        \textbf{s}_{2,j}\\
        ...
    \end{bmatrix}=
    \begin{bmatrix}
        a_{1,1} & e_{1,2} & b_{1,3}\\ 
        a_{2,1} & e_{2,2} & b_{2,3}\\ 
        ... & ... & ...
    \end{bmatrix}.
    \label{eq:shapley_matrix}
\end{equation}

To examine the contributions of the various parameters (i.e., ML model accuracy, energy consumption, network bandwidth) as well as conveniently distribute the reward under a budget for each node, the Shapley values are normalized so that they can be summed up to one. To do so, we use the following equation

\begin{align}
        \hat{R}_{|N|\times 3}^j
        &=R_{|N|\times 3}^j\times
        \begin{bmatrix}
            \frac{1}{\|\textbf{a}\|_1} & 0 & 0\\ 
            0 & \frac{1}{\|\textbf{e}\|_1} & 0\\ 
            0 & 0 & \frac{1}{\|\textbf{b}\|_1}
        \end{bmatrix} \nonumber \\
        &=\begin{bmatrix}
            a_{1,1} & e_{1,2} & b_{1,3}\\ 
            a_{2,1} & e_{2,2} & b_{2,3}\\ 
            ... & ... & ...
        \end{bmatrix}\times 
        \begin{bmatrix}
            \frac{1}{\|\textbf{a}\|_1} & 0 & 0\\ 
            0 & \frac{1}{\|\textbf{e}\|_1} & 0\\ 
            0 & 0 & \frac{1}{\|\textbf{b}\|_1}
        \end{bmatrix} \nonumber \\
        &=\begin{bmatrix}
            \frac{a_{1,1}}{\|\textbf{a}\|_1} & \frac{e_{1,2}}{\|\textbf{e}\|_1} & \frac{b_{1,3}}{\|\textbf{b}\|_1}\\ 
            \frac{a_{2,1}}{\|\textbf{a}\|_1} & \frac{e_{2,2}}{\|\textbf{e}\|_1} & \frac{b_{2,3}}{\|\textbf{b}\|_1}\\ 
            ... & ... & ...
        \end{bmatrix} \nonumber \\
        &=\begin{bmatrix}
            \hat{a}_{1,1} & \hat{e}_{1,2} & \hat{b}_{1,3}\\ 
            \hat{a}_{2,1} & \hat{e}_{2,2} & \hat{b}_{2,3}\\ 
            ... & ... & ...
        \end{bmatrix}
        \label{eq:nornalized_shapley_value_matrix}
\end{align}
\noindent to transform the original Shapley value matrix $R_{|N|\times 3}^j$ into a normalized Shapley value matrix $\hat{R}_{|N|\times 3}^j$. It multiples the Shapley value matrix with a diagonal matrix of the reciprocal of $l_1$ norm of each metric so that each element of the matrix can be divided by their $l_1$ norm and be normalized. 

To conveniently reward nodes based on their contribution, the three elements of the Shapley value vectors (i.e., accuracy, energy consumption, and network bandwidth) are combined into a single value. By doing so, we can reward them with a token of a single value rather than a vector. We use the following equation
\begin{align}
        \hat{\textbf{s}}_{|N|,j} 
        &=\hat{R}_{|N|\times 3}^j\times
        \begin{bmatrix}
            1\\ 
            1\\ 
            1
        \end{bmatrix}\times \frac{1}{3} \nonumber \\
        &=\begin{bmatrix}
            \hat{a}_{1,1} & \hat{e}_{1,2} & \hat{b}_{1,3}\\ 
            \hat{a}_{2,1} & \hat{e}_{2,2} & \hat{b}_{2,3}\\ 
            ... & ... & ...
        \end{bmatrix}\times
        \begin{bmatrix}
            1\\ 
            1\\ 
            1
        \end{bmatrix}\times \frac{1}{3} \nonumber \\
        &=\begin{bmatrix}
            \hat{s}_{1,j}\\ 
            \hat{s}_{2,j}\\ 
            ...
        \end{bmatrix}^T
    \label{eq:shapley_value_vector}
\end{align}
\noindent to combine them and compute the averaged sum of the three elements of the Shapley value vectors. The rationale behind this aggregation is as follows: suppose the budget for rewarding the training is one token, the budget for rewarding the energy saving is also one token, and the same for the network bandwidth. However, we eventually end up only having a total budget of one so the reward for training, energy saving, and network bandwidth should be divided by three to share the one token budget.

The above steps of computing the Shapley value vector $\hat{\textbf{s}}_{|N|,j}$ only happen in a local environment of the node $j$. Hence each node has its own local version, denoted as $\hat{\textbf{s}}_{|N|,*}$. To examine it conveniently we show it as a matrix
\begin{equation}
    \hat{G}_{|N|\times |N|}=
    \begin{bmatrix}
        \hat{\textbf{s}}_{|N|,1}\\ 
        \hat{\textbf{s}}_{|N|,2}\\ 
        ...
    \end{bmatrix}=
    \begin{bmatrix}
        \hat{s}_{1,1} & \hat{s}_{1,2} & ...\\ 
        \hat{s}_{2,1} & \hat{s}_{2,2} & ...\\ 
        ... & ... & ...
    \end{bmatrix},
    \label{eq:shapley_value_matrix_final}
\end{equation}
\noindent where $\hat{G}_{|N|\times |N|}$ is the Shapley value matrix stacked by each node's local $\hat{\textbf{s}}_{|N|,*}$, $\hat{s}_{i,j}$ is the Shapley value of the node $i$ evaluated by the node $j$. 

We need nodes to reach a consensus about $\textbf{s}$ so we sum them up and get the average version of the Shapley value, which is calculated as
\begin{equation}
    \hat{s}_i^*=\frac{1}{n}\sum_{j=0}^n \hat{G}_{i,j}=\frac{1}{n}\sum_{j=0}^n \hat{s}_{i,j}.
    \label{eq:shapley_value_matrix_average}
\end{equation}
Then, we get the final version of the Shapley value vector
\begin{equation}
    \textbf{s}=
    \begin{bmatrix}
        \hat{s}_1^*\\ 
        \hat{s}_2^*\\ 
        ...
    \end{bmatrix}^T,
\end{equation}
where each element in \textbf{s} is the contribution of the corresponding node.

\section{Experimental Evaluation}
\label{sec:evaluation}

We conducted experiments in a simulated environment, where AICons and other state-of-the-art consensus algorithms are simulated in similar way (i.e., same mining difficulty, same block size, same network delay, and same block broadcasting speed) to evaluate AICons's performance with metrics related to {\it scalability}, {\it reward}, and {\it fairness}. We will show the improvements of AICons by comparing it to four state-of-the-art consensus algorithms (i.e., PoW, PoS, PoD, PoFL). We also conducted an ablation study to evaluate the impact of the ablation of one or two of the three elements (i.e., ML model accuracy, energy consumption, and network bandwidth) of the Shapley value.

\subsection{Experiment Settings}
The experiments are simulated in a virtual private server (VPS) with Ubuntu 22.04 installed, 8-threaded CPU, 32 GB RAM and T4 GPU. The implementation is carried out by using PyTorch 1.13.1 (to implement AICons and ML tasks in PoFL) with CUDA 11.7 (to accelerate training and running neural networks), Numpy 1.24.1 (to store matrix data), scikit-learn 1.2.0 (to do polynomial linear regression) and Python 3.10.6. 

To train AICons, we collected data from Google Cluster Workload Traces \cite{reiss2011google} and Kaggle to construct the needed data. The data we prepared for the training contains 10000 records. The data attributes include CPU usage, memory usage, CPU cycles per instruction, and bandwidth. After preparing the data, we manually labelled winners from the data. AICons is implemented as four layers, where the first layer is the input layer, the second and third layers are hidden layers and the last layer is the output layer. This neural network design aims to reduce the training workload for a single VPS while maintaining a reasonable ML model accuracy.

To evaluate the performance of AICons regarding fairness, we used the  \textit{reward-contribution ratio} as a metric, which is computed by dividing the reward gained by each node by its contribution in terms of accuracy (only available for AICons and PoFL), energy consumption, and network bandwidth. A node who contributes more is supposed to be rewarded more, and vice versa. All nodes get rewards in proportion according to their contributions when all nodes get the same reward-contribution ratio, which means fair. In this paper, the contribution is measured by using the Shapley value. We used this metric since fairness is a common problem in various blockchain consensus algorithms such as PoW, PoUW~\cite{baldominos_coinai_2019, chenli_energy-recycling_2019, bravo-marquez_proof--learning_2019, lihu_proof_2020, lan_proof_2021, qu_proof_2021} where the contributions of non-winners are not rewarded.

To evaluate the performance of AICons regarding scalability, we adopted throughput as a metric that measures the number of transactions per second. The throughput is an important index in evaluating the scalability of blockchain systems since blockchain systems with low throughput are usually hard to be largely scaled and applied to various scenarios.

To study how the three elements of the Shapley value affect the total reward, we measure accuracy versus reward, energy consumption versus reward, and network bandwidth versus reward. This is important in evaluating how Shapley value measures the contribution of each node based on the combination of the three parameters (i.e., ML model accuracy, energy consumption, and network bandwidth).

To evaluate the performance of AICons regarding profitability, we measured how many rewards a node expects to earn when the number of nodes is increasing. If the number of rewards a node expects to earn decreases sharply when the number of nodes is increasing, the nodes are less profitable to join the blockchain system, resulting in a decreased scale of the blockchain system. The profit is computed by 
\begin{equation}
    \text{profit}=\text{reward} \times \text{price} - \text{cost},
    \label{eq:profit}
\end{equation}
where the price is the ETH market price. We set it to be 1798.68 AUD, which is based on the market price sampled on 06 January 2023 from \cite{noauthor_ether_nodate}. The cost is calculated by multiplying the CPU TDP, and the running time, and the electricity rates in Australia (accessed on 06 January 2023)~\cite{noauthor_victorian_nodate}.

To evaluate the improvements of AICons against the traditional consensus algorithms, we compared it with the four state-of-the-art consensus algorithms (discussed in Section~\ref{sec:related_work} i.e., PoW, PoS, PoD and PoFL) using the metrics previously discussed.

\subsection{Fairness}
\begin{figure}[http]
  \centering
  \includegraphics[width=.8\linewidth]{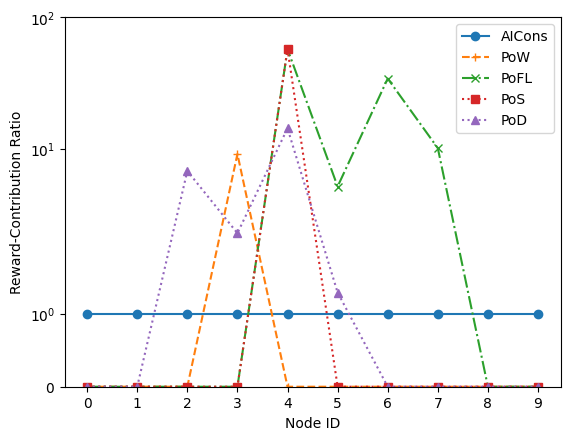}
  \caption{Fairness Analysis: Compare the reward-contribution ratio of different nodes when adopting different consensus algorithms on ten nodes in the blockchain network.}
  \label{fig:fairness}
\end{figure}

Figure~\ref{fig:fairness} shows the reward-contribution ratio of the ten nodes while they adopt different consensus algorithms. For AICons, each node gets a reward-contribution ratio of one since each node is rewarded based on its contribution. For PoW, only node \#3 gets a reward-contribution above zero since only the winner node gets the reward. For PoFL, only nodes \#4 -- \#7 get a reward-contribution ratio above zero while other nodes get a zero reward-contribution ratio since only a group of nodes whose global ML model has the highest accuracy are rewarded. For PoS, only node \#4 gets a reward-contribution ratio of around 60 since only node \#4  gets the reward while others are ignored. For PoD, only nodes \#2 -- \#5 get a reward-contribution ratio above zero since only a group of validators get rewarded. Above all, AICons is fair since each node is rewarded based on their contribution rather than being ignored.

\subsection{Ablation Study}
\begin{figure}[http]
  \centering
  \includegraphics[width=.8\linewidth]{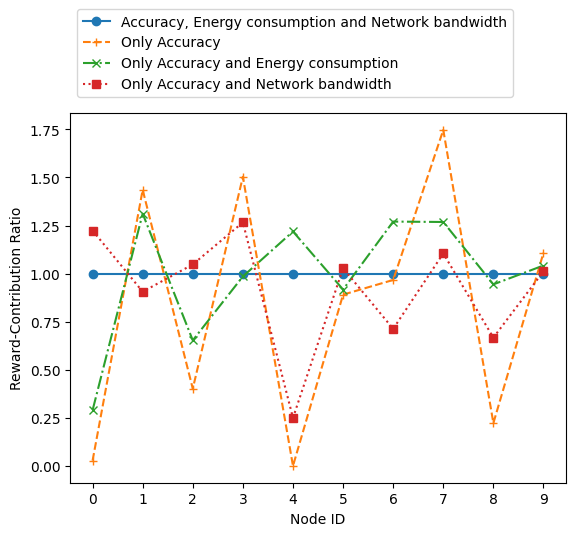}
  \caption{Ablation Analysis: Compare the reward-contribution ratio of different nodes when ablating one or two elements from the Shapley value.}
  \label{fig:ablation_lines}
\end{figure}

Figure~\ref{fig:ablation_lines} shows the reward-contribution ratio of ten nodes when ablating one or two elements from the Shapley value. All nodes get identically proportioned rewards according to their contributions when all nodes get the same reward-contribution ratio, which means fair.

When there is no ablation, the reward-contribution ratio is one for each node since they are rewarded based on their contribution with a 1:1 ratio. When only accuracy is considered, the reward-contribution ratio fluctuates between 0 and 1.75. Node zero gets a reward-contribution ratio of around 0.03 since it has a low ML model accuracy, and the contribution of energy consumption, as well as network bandwidth, is weakened. Node \#7 has a high reward-contribution ratio since its ML model accuracy is high. However, its energy consumption and network bandwidth contribution are not considered as a negative impact. When only accuracy and energy consumption are considered, the reward-contribution ratio fluctuates between 0.3 and 1.3. When only accuracy and network bandwidth are considered, the reward-contribution ratio fluctuates between 0.2 and 1.2. 

The experimental results in Figure~\ref{fig:ablation_lines} show that the more parameters are considered in the Shapley value, the less degree of fluctuation we have. It also means that AICons is closer to being fair. When all the three parameters are  considered, all the works of accuracy, energy consumption, and network bandwidth are rewarded accordingly and fairness is maintained.

\subsection{Scalability}

\begin{figure}[http]
  \centering
  \includegraphics[width=.8\linewidth]{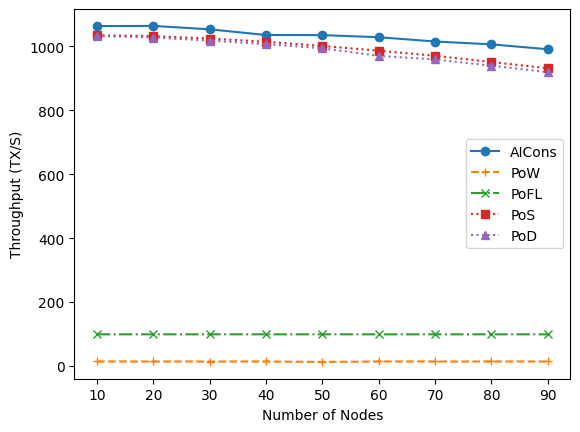}
  \caption{Scalability Analysis: Compare the throughput of AICons with other schemes when increasing the number of nodes in the blockchain network.}
  \label{fig:scalability}
\end{figure}

Figure~\ref{fig:scalability} shows how throughput changes when the number of nodes is increasing. AICons has the highest throughput since the neural network inference is fast when GPU computing is available (i.e., GPU acceleration). When the number of nodes increases, its throughput decreases slightly because the trained ML model still can recommend a winner in a short time even though there are more nodes in the blockchain. PoW has the lowest throughput since its mining difficulty limited the throughput and the block size is fixed so the throughput does not change when the number of nodes is increasing. PoFL also has poor throughput since it takes a long time to train a federated ML model in each consensus round. PoS and PoD have a high throughput since they use a node selection algorithm to select nodes. However, the throughput decreases when the number of nodes increases since the node selection algorithm consumes more time when the number of nodes is larger. Above all, AICons has an advantage in using GPU computing to help it scale to a large blockchain network.

\subsection{Rewards}

\subsubsection{ML accuracy vs Rewards}

\begin{figure}[http]
  \centering
  \includegraphics[width=.8\linewidth]{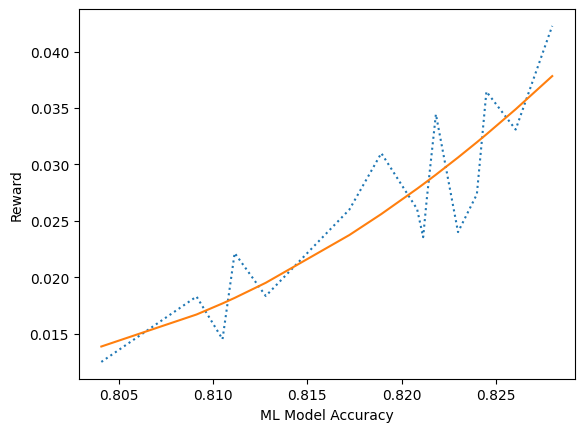}
      \caption{AICons issues more rewards to nodes that contribute ML models with higher accuracy. The dotted blue line is the actual reward. The solid orange line is polynomial linear regression.}
  \label{fig:security}
\end{figure}

Figure~\ref{fig:security} shows the rewards received by nodes versus their ML model accuracy. When a node submits a poor ML model and lowers the global model accuracy, it receives less reward as punishment. According to Figure~\ref{fig:security}, when the ML model accuracy of a node increases, the reward is increasing accordingly. The dotted blue line is the actual reward. The fluctuation is caused by other factors except for ML model accuracy such as energy consumption and network bandwidth. To smooth the line, we used a polynomial linear regression (i.e., the orange solid line) to approach the dotted blue line to show the trend of increased rewards caused by increased accuracy. Since the low ML model accuracy means less reward, AICons punishes those who submit poor quality ML models with less reward.

\subsubsection{Energy consumption vs Rewards}
\begin{figure}[http]
  \centering
  \includegraphics[width=.8\linewidth]{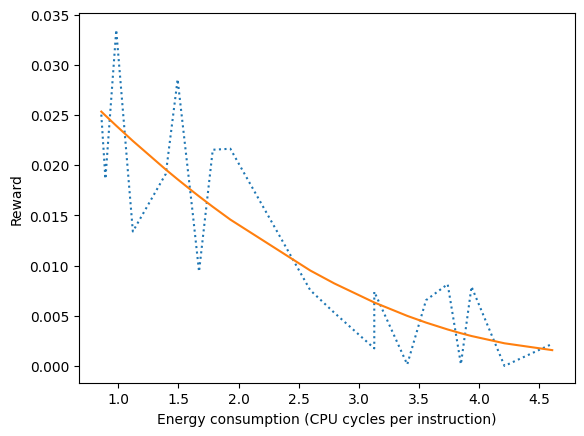}
  \caption{AICons issues fewer rewards to nodes that consume more energy. The dotted blue line is the actual reward. The solid orange line is polynomial linear regression.}
  \label{fig:security_energy}
\end{figure}

Figure~\ref{fig:security_energy} shows the rewards received by nodes versus their energy consumption. If a node consumes lots of energy to get higher accuracy and higher reward, it leads to a computing power competition, which is not energy efficient. Besides, it could cause cheating behaviour of nodes, such as consuming more energy to over-train the ML model. To punish energy waste and mitigate cheating behaviours, we gave fewer rewards to those who consumed lots of energy. According to Figure~\ref{fig:security_energy}, when energy consumption decreases, the reward is increasing, and vice versa. This demonstrates that in AICons, the node who consumes more energy will be punished by being assigned less reward. 

\subsubsection{Network bandwidth vs Rewards}
\begin{figure}[http]
  \centering
  \includegraphics[width=.8\linewidth]{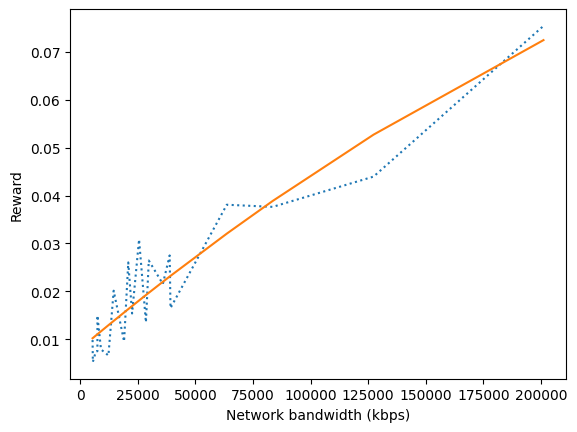}
  \caption{AICons issues more rewards to nodes that provide more network bandwidth. The dotted blue line is the actual reward. The solid orange line is polynomial linear regression.}
  \label{fig:security_bandwidth}
\end{figure}

Figure~\ref{fig:security_bandwidth} shows the rewards received by nodes versus their network bandwidth. A node who contributes less bandwidth to a blockchain system will increase delays in transmitting data, affecting the scalability of the blockchain system. To punish this behaviour, we gave less reward to those who contribute less network bandwidth. According to Figure~\ref{fig:security_bandwidth}, when the network bandwidth increases, the reward is also increasing, which means those who contribute less bandwidth to a blockchain system are punished. In Figure~\ref{fig:security_bandwidth}, the reward fluctuates when the network bandwidth is between 5000 kbps and 50000 kbps and it is smooth when the network bandwidth is above 75000 kbps. This is because the network bandwidth distribution of above 75000 kbps is rare in the dataset.

\subsection{Profit}
\begin{figure}[http]
  \centering
  \includegraphics[width=.8\linewidth]{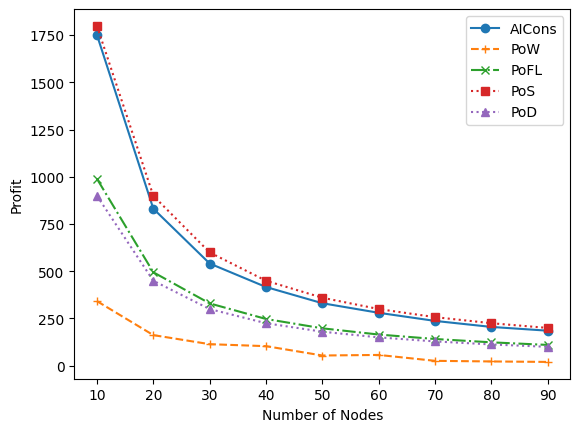}
  \caption{The impact of the numbers of participants on profit}
  \label{fig:termination}
\end{figure}

Figure~\ref{fig:termination} shows when the number of nodes in the blockchain network increases, the average profit of each node is decreasing. This is due to that more nodes compete for a single unit of profit. In the experiment, AICons and PoS are the most profitable consensus algorithms. PoW is the least profitable consensus algorithm. PoFL and PoD are neutral in profitability. AICons is more profitable compared with PoW, PoFL, and PoD, because it consumes less energy. PoD also consumes less energy but the rewards are divided by more validators. Although PoS is slightly more profitable than AICons, AICons has better fairness than PoS when it comes to the reward distribution based on contribution.

\section{Conclusion}
\label{sec:conclusion}

This paper described an AI-enabled consensus algorithm, named AICons, for blockchain systems to effectively generate blocks, along with a fair rewarding mechanism that encourages nodes to reduce energy consumption. In particular, we innovatively utilised decentralised federated learning to train a recommender ML model for selecting winners. Future work will include exploring the possibility to use an unsupervised ML model to recommend the winner, which simplifies the step of manual data labelling and mitigates the preference of certain winners.

\bibliographystyle{ieeetr}
\bibliography{main.bib}

\end{document}